# Throughput Scaling for MMF-Enabled Optical Datacenter Networks by Time-Slicing-Based Crosstalk Mitigation

Zhizhen Zhong[1,2], Nan Hua[1,2,*], Yufang Yu[1,2], Zhongying Wu[3], Juhao Li[3], Haozhe Yan[1,2], Shangyuan Li[1,2], Ruijie Luo[1,2], Jialong Li[1,2], Yanhe Li[1,2], and Xiaoping Zheng[1,2,*]

1. Tsinghua National Laboratory for Information Science and Technology (TNList), Tsinghua University, Beijing 100084, China.
2. Department of Electronic Engineering, Tsinghua University, Beijing 100084, China
3. State Key Laboratory of Advanced Optical Communication Systems and Networks, Peking University, Beijing 100871, China.

*Correspondence: huan@tsinghua.edu.cn, xpzheng@mail.tsinghua.edu.cn

**Abstract:** Modal crosstalk is the main bottleneck in MMF-enabled optical datacenter networks with direct detection. A novel time-slicing-based crosstalk-mitigated MDM scheme is first proposed, then theoretically analyzed and experimentally demonstrated.

## 1. Introduction

Datacenter is becoming the basic infrastructure for future information-based society. More and more datacenters are being constructed to meet the increasing user demands. Meanwhile, datacenters are also evolved to be larger and larger with thousands of racks and complex interconnections. Optical fibers, as the key enabler for efficient intra-datacenter networks, also need to adapt to these new challenges. Conventional Single-Mode Fiber (SMF) can hardly deal with such traffic surge within limited financial budget. Therefore, short-reach Multi-Mode Fiber (MMF) [1] with direct detection is regarded as a promising candidate for intra-datacenter optical interconnect networks, due to its extremely large capacity and low cost over SMF.

However, MMF severely suffers from modal crosstalk in direct-detection-based short-reach networks. Such modal crosstalk accumulate along propagation path, and can induce OSNR degradation at the receiver end [2,3]. In some cases, the OSNR degeneration can result in signal transmission failure, so that relating traffic requests cannot be accepted by the network. Such crosstalk must be prevented, as it cannot be fully undone by electrical signal processing after direction detection [4]. Therefore, this bottleneck restricts the scaling of network throughput as the upgrade from SMF to MMF, leading to networks' resource inefficiency and operators' revenue loss.

To solve this bottleneck, we propose a time-slicing-based crosstalk-mitigated Mode Division Multiplex (MDM) scheme in direct-detection short-reach MMF networks. A theoretical analysis is performed to acquire its optimal performance. And an experimental demonstration is carried out to verify its feasibility in MMF networks.

## 2. Time-slicing-based crosstalk-mitigated MDM scheme

Fig. 1 depicts the principle of time-slicing-based crosstalk mitigation scheme enabled by Optical Time Slice Switching (OTSS) (OTSS principle in Fig. 1(a), for more information about OTSS, refer to [5]). In conventional MDM transmission (Fig. 1(b)), multiple modes are used for transmission at the same time, resulting in severe modal crosstalk that affects signal OSNR at the receiver side. But, our scheme (Fig. 1(c)) can achieve less, or no crosstalk by avoiding using modes with high crosstalk in the same time slice, while changing utilized modes in different time slices by switching at selected switching points, assisted by precise synchronized time [6]. The basic idea for designing time-slicing-based crosstalk mitigation is to stagger utilized modes in temporal domain via synchronized time slices to alleviate or avoid the severe modal crosstalk, thus increasing network throughput.

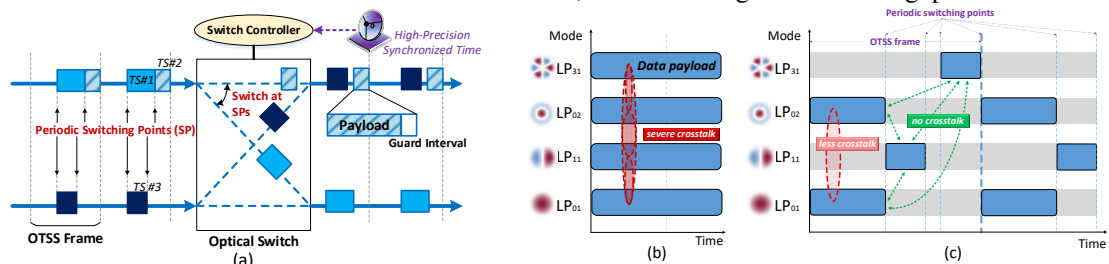

Fig. 1, (a) OTSS principle at a node, (b) conventional MDM transmission, (c) time-slicing-based crosstalk-mitigated MDM.

## 3. Theoretical analysis

In MDM systems, modal crosstalk is dependent on various factors, such as fiber fabrication, imperfections, bending or twisting, etc. [4]. Accumulated crosstalk ($XT$) can be described as: $XT = \tanh(hz)$, where $h$ is the mode coupling parameter (m$^{-1}$) and $z$ is the fiber length (m) [7]. If $z$ is in logarithmic scale, and $XT$ is measured in dB, $XT$ can be approximated to be linear to z. So, $XT \sim \log(z)$. In intra-datacenter networks, fiber links are usually hundreds of meters. The time slice propagation delay (~μs) is relative negligible when time slices are in ~100 μs scale. Therefore, propagation delay can be omitted. Finally, we formulate the problem into a Mixed Integer Program (MIP).

*This work was supported in part by projects under National 973 Program grant No. 2014CB340104/05, and NSFC under grants No. 61435006, 61427813, 61621064.*

**Parameters:**
- $G(N, E)$: network topology in a unidirectional graph, where $N$ and $E$ denotes the set of nodes and MMF fiber links, respectively.
- $R$: set of traffic requests.
- $s_r, d_r, b_r$: source, destination, and bandwidth of traffic request $r$, $r \in R$.
- $X$: accumulated crosstalk threshold for direct-detection receivers.
- $M$: set of supporting modes by a MMF.
- $d(i,j)$: length of fiber link $(i,j)$ of fiber link, $(i,j) \in E$.
- $C$: maximum link capacity.
- $Y(m_1, m_2)$: modal crosstalk of mode $m_1, m_2$, in dB.
- $Max$: a maximum number.
- $\eta_1, \eta_2$: parameters for optimization sequence, $\eta_1 \gg \eta_2$.
- $T$: OTSS frame length.
- $S$: length of the minimum time slice.

**Variables:**
- $\lambda_{(i,j)}^{r,m,t}$: binary, which equals one if request $r$ uses mode $m$ and time slot $t$ on fiber link $(i,j)$.
- $\beta_{(i,j)}^{r_1,r_2,m_1,m_2,t}$: binary, which equals one if request $r_1$ on mode $m_1$, $r_2$ on mode $m_2$ have crosstalk on time slot $t$ of fiber link $(i,j)$.
- $\theta_{(i,j)}^{r_1,r_2,m_1,m_2}$: binary, which equals one if request $r_1$ on mode $m_1$, $r_2$ on mode $m_2$ have crosstalk on fiber link $(i,j)$.
- $\rho_r$: binary, which equals one if request $r$ is accepted.

**Objective:** Maximize network throughput first, then minimize network resource usage.

$$\text{Maximize}: \eta_1 \cdot \sum_{r \in R} \rho_r \cdot b_r - \eta_2 \sum_{r \in R} \sum_{m \in M} \sum_{t \in T} \sum_{(i,j) \in E} \lambda_{(i,j)}^{r,m,t} \quad (1)$$

**Constraints:**

$$\sum_{j \in N} \sum_{m \in M} \sum_{t \in T} \lambda_{(i,j)}^{r,m,t} - \sum_{j \in N} \sum_{m \in M} \sum_{t \in T} \lambda_{(j,i)}^{r,m,t} = \begin{cases} \rho_r \cdot b_r, i = s_r \\ -\rho_r \cdot b_r, i = d_r, \quad \forall r \in R \\ 0, i \neq s_r, d_r \end{cases} \quad (2)$$

$$\sum_{m \in M} \sum_{j \in N} \lambda_{(s_r,j)}^{r,m,t} = \sum_{m \in M} \sum_{i \in N} \lambda_{(i,d_r)}^{r,m,t}, \forall r \in R, t \in T \quad (3)$$

$$\sum_{m \in M} \sum_{j \in N} \lambda_{(z,j)}^{r,m,t} = \sum_{m \in M} \sum_{i \in N} \lambda_{(i,z)}^{r,m,t}, \forall r \in R, t \in T, z \in N \setminus \{s_r, d_r\} \quad (4)$$

$$\sum_{m \in M} \sum_{j \in N} m \cdot \lambda_{(s_r,j)}^{r,m,t} = \sum_{m \in M} \sum_{i \in N} m \cdot \lambda_{(i,d_r)}^{r,m,t}, \forall r \in R, t \in T \quad (5)$$

$$\sum_{m \in M} \sum_{j \in N} m \cdot \lambda_{(z,j)}^{r,m,t} = \sum_{m \in M} \sum_{i \in N} m \cdot \lambda_{(i,z)}^{r,m,t}, \forall r \in R, t \in T, z \in N \setminus \{s_r, d_r\} \quad (6)$$

$$\sum_{r \in R} \lambda_{(i,j)}^{r,m,t} \leq 1, \quad \forall m \in M, t \in T, (i,j) \in E \quad (7)$$

$$\sum_{t \in T} (\langle \lambda_{(i,j)}^{r,m,t} \neq \lambda_{(i,j)}^{r,m,(t+1)\%N_m} \rangle \times 1) \leq 2, \quad \forall r \in R, m \in M, (i,j) \in E \quad (8)$$

$$\sum_{t \in T} (\langle \sum_{m \in M} \lambda_{(i,j)}^{r,m,t} \neq \sum_{m \in M} \lambda_{(i,j)}^{r,m,t+1} \rangle \times 1) \leq 2, \quad \forall r \in R, (i,j) \in E \quad (9)$$

$$\sum_{m' \in M} \sum_{t' \in T} \lambda_{(i,j)}^{r,m',t'} - (\lambda_{(i,j)}^{r,m,t} - 1) \cdot Max \cdot C \cdot S/T \geq b_r, \forall r \in R, m \in M, t \in T, (i,j) \in E \quad (10)$$

$$\sum_{r_2 \in R} \sum_{(i,j) \in E} \sum_{m_1 \in M} \sum_{m_2 \in M} \theta_{(i,j)}^{r_1,r_2,m_1,m_2} \cdot d(i,j) \cdot Y(m_2, m_1) \leq X, \quad \forall r_1 \in R \quad (11)$$

$$\sum_{t \in T} \beta_{(i,j)}^{r_1,r_2,m_1,m_2,t}/Max \leq \theta_{(i,j)}^{r_1,r_2,m_1,m_2} \leq \sum_{t \in T} \beta_{(i,j)}^{r_1,r_2,m_1,m_2,t}, \forall r_1, r_2 \in R, m_1, m_2 \in M, (i,j) \in E \quad (12)$$

$$\beta_{(i,j)}^{r_1,r_2,m_1,m_2,t} \geq \lambda_{(i,j)}^{r_1,m_1,t} + \lambda_{(i,j)}^{r_2,m_2,t} - 1, \forall r_1, r_2 \in R, m_1, m_2 \in M, t \in T, (i,j) \in E \quad (13)$$

$$\beta_{(i,j)}^{r_1,r_2,m_1,m_2,t} \leq \lambda_{(i,j)}^{r_1,m_1,t}, \forall r_1, r_2 \in R, m_1, m_2 \in M, t \in T, (i,j) \in E \quad (14)$$

$$\beta_{(i,j)}^{r_1,r_2,m_1,m_2,t} \leq \lambda_{(i,j)}^{r_2,m_2,t}, \forall r_1, r_2 \in R, m_1, m_2 \in M, t \in T, (i,j) \in E \quad (15)$$

Equation (Eq.) (2) is the multi-commodity flow-conservation equation for request routing. Eqs.(3-4)/(5-6) ensure time slice/mode continuity on different links along the routing path. Eq.(7) ensures a time slice is used once. Eq.(8) ensures time slices used by a request should be contiguous on a link. Eq.(9) ensures used time slices should be the same on different modes. Eq.(10) ensures the capacity of used time slices is enough for the request. Eq.(11) ensures a request's accumulated crosstalk cannot exceed the threshold. Eqs.(12-15) define relationships among variables. The operator $\langle \cdot \rangle \times 1$ represents the logical calculation that returns one if the expression inside the angle brackets is true.

We solve the MIP via a commercial IBM CPLEX platform. The threshold for accumulated crosstalk is -13dB under direct detection. Crosstalk between selected modes are shown in Tab. 1. In fact, as modal crosstalk can vary among different fibers, and between modes with different coupling strength, these selected modes can be mapped to any possible modes in certain circumstances. We adopt small-scale fat-tree datacenter topology as shown in Fig. 2. Each fiber is 100 m length for general datacenter size. Traffic requests are generated uniformly between edge switch pairs with bandwidth incremental granularity of 1Gb/s. The frame length of OTSS is set to be 20 ms, and the minimum time slice is 5 ms. Each modal channel has a capacity of 10 Gb/s. Only 1550 nm wavelength is adopted.

Tab. 1, crosstalk (dB/100m) of selected modes.

| XT | $m_1$ | $m_2$ | $m_3$ | $m_4$ |
|---|---|---|---|---|
| $m_1$ | - | -26.0 | -21.2 | -43.0 |
| $m_2$ | -17.7 | - | -15.8 | -19.7 |
| $m_3$ | -19.5 | -14.3 | - | -15.6 |
| $m_4$ | -21.5 | -16.7 | -17.5 | - |

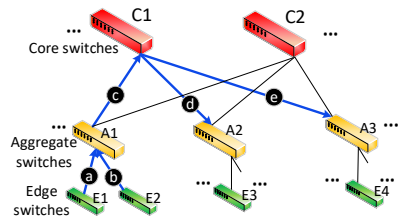

Fig. 2, a fat-tree datacenter topology.

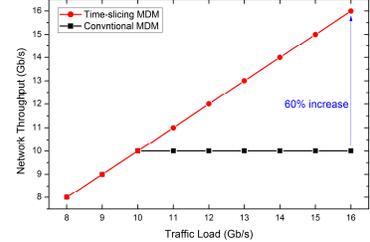

Fig. 3, network throughput vs. offered load

Fig. 3 depicts the relationship between network throughput and traffic load. We can find that our time-slicing-based MDM scheme can achieve at least 60% higher throughput than conventional MDM networks under heavy-loaded traffic. It should be pointed that this throughput scaling comes from flexible allocation of time slices, this increase will be more significant when the number of adopted wavelengths and modes become larger.

## 4. Experimental demonstration

We also conduct an experimental demonstration to verify the feasibility of time-slicing-based crosstalk-mitigated MDM scheme. Fig. 3 shows the experimental setup based on a 4-mode MDM transmission systems of OM3 MMF with core/cladding diameter of 50/125 μm [8]. Xilinx VC709 FPGA is used to generate $2^7-1$ Pseudo-Random Binary Sequence (PRBS) codes as signal source for optical transmitters (SFP+) at 10Gb/s carried on 1550 nm wavelength, and also analyze signal symbols inside the time slice at the receiver. In this experiment, the minimum time slice is set to be 500 μs within 20 ms frame length, and guard interval is 50 μs. Two Magneto-electronic (MO) $2 \times 2$ optical switches (SW) are used for time slice switching periodically at precise synchronized time under OTSS principle [5].

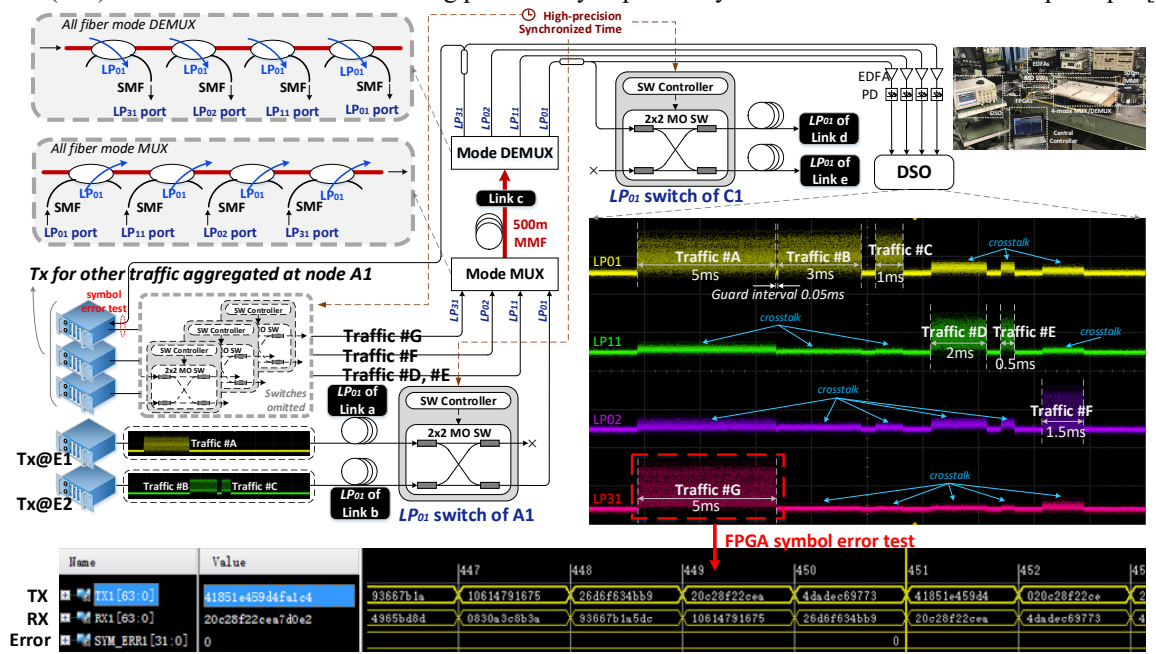

Fig. 4, experimental setup and results.

Fig. 4(a) shows the experimental setup and results. We select 5 links (Link a,b,c,d,e) on the topology of Fig. 2 to demonstrate our scheme. Signals on four mode of Link c is shown to observe the distribution of times slices and corresponding crosstalk. We find that Traffic #A from Node E1, Traffic #B and #C from Node E2 are aggregated at Node A1 on mode $LP_{01}$, and are both transmitted on Link c. There are also four more traffic requests (Traffic #D, #E, #F, #G) aggregated at Node A1 on different modes ($LP_{11}$, $LP_{02}$, $LP_{31}$). These signals induce noticeable crosstalk among different modes after 500m propagation on Link c, as the DSO screenshot shows. Thanks to our time-slicing-based crosstalk-mitigated MDM scheme, Traffic #G on $LP_{31}$ only suffer crosstalk from Traffic #A on $LP_{01}$. Such crosstalk do not affect the signal quality of Traffic #G, which can still achieve zero symbol error after 500m propagation over MMF. However, if four or more modes co-propagate, severe crosstalk may result in high symbol error rate, or even service outages. By allocating time slices wisely, we can mitigate modal crosstalk, increase traffic acceptance and scale network throughput.

## 5. Conclusion

In this paper, a time-slicing-based crosstalk-mitigated MDM scheme is proposed for mitigating modal crosstalk in MMF datacenter networks with direction detection. Theoretical analysis show that over 60% throughput increase can be achieved, and experimental demonstration proves the feasibility of the proposed scheme.

## 6. References

[1] G. Li, *et al.*, "Space-division multiplexing: the next frontier in optical communication," *Advances in Optics and Photonics*, 2014.
[2] B. Franz, "Mode Group Division Multiplexing in Graded-Index Multimode Fibers," *Bell Labs Technical Journal*, 2013.
[3] F. Yaman, *et al.*, "Impact of Modal Crosstalk and Multi-Path Interference on Few-Mode Fiber Transmission," *Proc. OFC*, 2012.
[4] K.-P. Ho, *et al.*, "Linear propagation effects in mode-division multiplexing systems," *IEEE/OSA J. Lightw. Technol.*, 2014.
[5] N. Hua and X. Zheng, "All-optical time slice switching method and system based on time synchronization," *US Patent*, US 2016/0036555A1.
[6] A. S. Muqaddas, *et al.*, "Exploiting Time-Synchronized Operations in Software-defined Elastic Optical Networks," *Proc. OFC*, 2017.
[7] F. M. Ferreira, *et al.*, "Semi-Analytical Modelling of Linear Mode Coupling in Few-Mode Fibers," *IEEE/OSA J. Lightw. Technol.*, 2017.
[8] Z. Wu, *et al.*, "4-Mode MDM transmission over MMF with direct detection enabled by cascaded mode-selective couplers," *Proc. OFC*, 2017.